# Dynamic Optical Path Provisioning for Alien Access Links: Architecture, Demonstration, and Challenges

Hideki Nishizawa, Takeo Sasai, Takeru Inoue, Kazuya Anazawa, Toru Mano, Kei Kitamura, Yoshiaki Sone, Tetsuro Inui, and Koichi Takasugi

NTT Network Innovation Laboratories

*Abstract*— With the spread of Data Center Interconnect (DCI) and local 5G, there is a growing need for dynamically established connections between customer locations through high-capacity optical links. However, link parameters such as signal power profile and amplifier gains are often unknown and have to be measured by experts, preventing dynamic path provisioning due to the time-consuming manual measurements. Although several techniques for estimating the unknown parameters of such *alien access links* have been proposed, no work has presented architecture and protocol that drive the estimation techniques to establish an optical path between the customer locations. Our study aims to automatically connect customer-owned transceivers via alien access links with optimal quality of transmission (QoT). We first propose an architecture and protocol for cooperative optical path design between a customer and carrier, utilizing a state-of-the-art technique for estimating link parameters. We then implement the proposed protocol in a software-defined network (SDN) controller and white-box transponders using an open API. The experiments demonstrate that the optical path is dynamically established via alien access links in 137 seconds from the transceiver's cold start. Lastly, we discuss the QoT accuracy obtained with this method and the remaining issues.

## I. Introduction

DIGITAL coherent optical transmission technology was developed as a long-haul optical transmission technology. In the 2010s, with the advancement of silicon microfabrication technology, it has rapidly become more power efficient and modularized (with prices dropping over 25%/year) and is now implemented in various use cases other than long-haul. In particular, the Data Center Interconnect (DCI) market is rapidly expanding at a CAGR of more than 10%. In the local 5G market, where the CAGR is approaching 50%, there will be a rapidly growing demand for optical networks to connect local 5G sites that are located far from each other. To meet the emerging demands for DCI and local 5G, optical paths should occasionally be rebuilt. Aside from DCI and local 5G, in an emergency such as a disaster, the network must be re-established by finding an appropriate transceiver configuration for the damaged fiber in a short period of time and conducting provisioning/restoration. In public networks, if low margin operation [1] becomes regular practice for effective use of optical resources, it is necessary to periodically update the transceiver configuration and routes before the quality of transmission (QoT) degrades. In addition, for low latency and power savings, it is desirable to directly connect customers with optical paths without optical-electrical-optical conversion whenever possible.

There are several challenges to dynamically setting up an optical path in practice. Optical paths often need to be set up for transmission links whose components are unknown or whose parameters must be re-measured. Coherent transceivers take only a few minutes to stabilize optical signals from a cold start, but it takes at least a few days for a carrier to set up an optical path in current operations. This is because experts take the time to investigate the link parameters such as fiber types, the optical signal power profile along the link, and amplifier gains when setting up an optical transmission link. Our goal is to automatically configure an optical path with optimal transmission quality over a transmission link whose current parameters are unknown. In this paper, we define an alien access link (AAL) as an access network whose components, quality, and parameters are unknown, and consider a situation in which customer transceivers are connected to a carrier-managed network with the AALs (Figure 1).

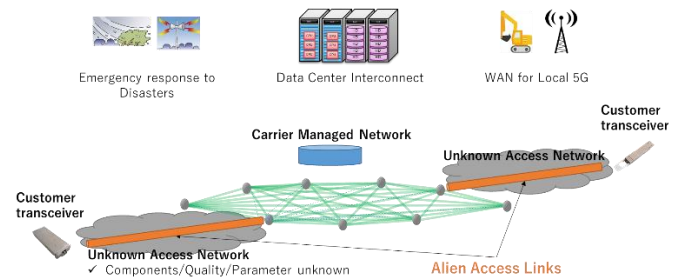

**Fig. 1.** Alien access links directly connecting consumers.

There have been several studies related to dynamic path provisioning via AALs. Advancements in digital signal processing and machine learning technologies have made it possible for receivers to estimate the QoT. The following two approaches have been proposed.

- Digital longitudinal monitoring (DLM): Estimates link parameters in a distance-wise and component-wise manner such as signal power profile, fiber types, gain spectra of each amplifier, etc. [2, 3] The estimated parameters are then used by an optical path design tool [4] to estimate the QoT.
- Black box approach: A characterized probing light with a fixed modulation format and symbol rate is inserted into the network, and the forward error correction (FEC) BER estimation of the receiver is used to estimate the end-to-end

(EtE) QoT. [5, 6, 7]

After QoT is estimated using these approaches, the optimal configuration can be set to the transceiver using an Open API such as the Transponder Abstraction Interface (TAI) defined in the Telecom Infra Project (TIP). If necessary, the line system can also be controlled by the OpenROADM YANG model or a similar model. Optical signals coming from AALs are likely to be alien wavelengths, which have been extensively researched [8]. Research is also being conducted to establish optical paths beyond administrative boundaries, such as between customers and carriers (GMPLS [9], alien wavelength [10]). Thus, although these elemental technologies exist, architectures and protocols needed for the use cases described at the beginning of this paper have not been studied.

The contributions of this paper are as follows:

- We compare the two QoT estimation approaches assuming that the transmission link between the carrier and the customer is unknown. After clarifying the advantages and disadvantages of the two approaches, the remainder of this paper focuses on the DLM approach because of its distance-based resolution which is indispensable to fault diagnosis.
- We propose an architecture and protocol based on the DLM approach. A key architectural issue is whether to place required functions, e.g., a DLM estimator, on a customer or carrier. The protocol is designed to exchange only disclosable information between a customer and carrier.
- Using the TIP's TAI, we implement the proposed protocol in an SDN controller and a white-box transponder. We demonstrate that the optical path can be configured from a cold start without a significant temporal overhead. We also discuss the accuracy of the resulting QoT based on the required margins.

Lastly, remaining research issues are discussed in detail, including those specific to practical operation.

## II. NETWORK MODEL AND REQUIREMENTS

Our study is based on the following simple network model (configuration and assumptions). We assume the configuration shown in Figure 1, in which the path traverses the customer transceiver, AAL, carrier-managed network, AAL, and customer transceiver. The customer transceivers and the carrier's line system may be produced by different vendors. The following assumptions are made for each in this paper.

- AALs: For simplicity, only a single wavelength channel is used in this paper (however, most of the arguments in this paper are valid even if WDM is assumed). We assume that neither the customer nor the carrier knows the link parameters.
- Carrier-managed network: Any optical device (amplifier, filter, etc.) may be installed. The carrier knows the link parameters, while the customers do not.
- Customer transceivers: We assume coherent transceivers. Only the customers know the transceiver characteristics, i.e., the optical signal to noise ratio at the transceiver output (TxOSNR), back-to-back bit error rate vs optical signal to noise ratio (BER vs OSNR), and the implemented functions. The route and wavelength used are given in advance, and the method for selecting them is out of scope. A secure channel is pre-constructed between the customers and the carrier, through which control information can be exchanged. Authentication is outside the scope of this paper.

Here, we define the requirements for optical path configuration as follows. The transmitting customer sends an optical path setup request to the receiving customer. Our goal is to maximize the performance on an optical path through an AAL. The performance is defined as the bit rate per optical frequency range. The allowable BER was assumed to be the standard threshold, i.e., $1 \times 10^{-12}$. The carrier and customers cooperatively configure an optimal optical path by calculating link OSNR degradation caused by amplified spontaneous emission noise (OSNR_ase) and fiber nonlinear noise (OSNR_nli). The customers configure the transceiver parameters (output power, bit rate, modulation format, FEC, etc.) for maximum link performance while taking into account the receiver sensibility. The configuration process is done automatically, without human intervention, and should take at most 10 minutes to complete.

## III. OPTICAL PATH DESIGN OVER AAL: A REVIEW OF EXISTING APPROACHES

This section presents two existing approaches that can be used for optical path design over AALs and discusses the advantages and disadvantages of each approach.

### A. DLM: Optical path design based on link parameter estimation of AALs

We use the DLM technique [2] to estimate the parameters of AALs. DLM estimates physical characteristics distributed in the fiber-longitudinal direction by solely processing the received (Rx) signals without traditional analog testing instruments (OTDR, OSA, etc.). For instance, the estimation of the fiber-longitudinal power profile (PP), gain spectra of optical amplifiers (OAs), and frequency responses of optical filters have been successfully demonstrated. Unlike analog devices, DLM is capable of "multi-span" measurement, which enables rapid and cost-efficient monitoring of the AAL. In addition, DLM visualizes the entire line system, making it possible to optimize and identify/localize soft-failures caused in link components such as fibers, amplifiers, and filters. In this work, we focus on the PP estimation (PPE).

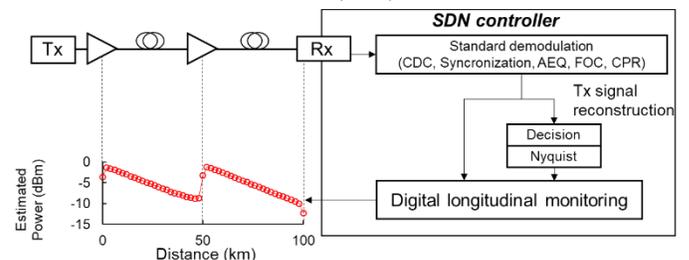

**Fig. 2.** Digital longitudinal monitoring.

Figure 2 shows a common configuration of DLM. The key to PP E is the fiber nonlinearity (self-phase modulation); the PP is obtained by estimating the distance-wise nonlinear phase rotation due to their proportionality. An arbitrary signal emitted from the transmitter (Tx) propagates an AAL, and the received signal is then sent to the SDN controller, in which the distance-wise nonlinearity is estimated through DSP. Note that no special training or pilot signal is needed for the Tx signal because it can be recovered at the receiver or SDN controller. To estimate the fiber nonlinearity caused in the link, transmission impairments except the nonlinearity are compensated for. The demodulated signals and the Tx signal recovered from symbol decision (decoding and mapping if possible) and Nyquist filtering are fed into the DLM algorithms. The distance-wise nonlinear phase rotation is obtained by performing least squares [2] or cross-correlation [11].

In this approach, the optical path is established in the following steps:
1) The Tx customer requests the carrier to establish an optical path.
2) The customer sends its transceiver characteristics to the carrier via the secure channel.
3) The customer and carrier synchronously estimate AAL parameters.
4) The customer or carrier selects the transceiver mode using the estimated link parameters.
5) The customer configures the transceiver with the selected mode.

The advantage of this approach is that, by inputting the estimated PP to existing optical path design tools, the main factors that degrade the line system (OSNR_ase, OSNR_nli) can be calculated. However, this approach also presents the following two issues: 1) To calculate the OSNR value at the receiver end, the TxOSNR value is needed in addition to the line system's estimated OSNR value; and 2) The analysis of the received BER requires the receiver's back-to-back characteristics such as BER vs. OSNR, in addition to the line system QoT. As described in Section IV, the proposed architecture and protocols enable carriers and customers to work together to exchange the TxOSNR and back-to-back characteristics, so the two issues can be resolved. This enables more accurate path designs and reduces redundant margin as described in Section V. Notably, unlike analog devices, DLM is capable of "multi-span" measurement, which enables rapid and cost-efficient monitoring of the AAL. In addition, DLM visualizes the entire line system, making it possible to optimize and identify/localize soft-failures caused in link components such as fibers, amplifiers, and filters. However, because DLM uses the fiber nonlinearity to estimate the above characteristics, their accuracy degrades at a practical launch power [2, 11, 12].

*B. Black box approach: Optical path design based on EtE QoT estimation with receiver pre-FEC BER*

The black box approach selects the optimal transceiver parameters by EtE QoT estimation with probing light without estimating the parameters of the AAL [5, 6, 7]. In this approach, the optical path is only set up by the customers, with no involvement from the carrier except to prepare an empty channel. The black box approach roughly follows these steps:

1) The Tx customer requests the carrier to prepare an empty channel.
2) The carrier prepares the channel and notifies the customer of its wavelength.
3) The customer configures transceivers with the wavelength.
4) The customer selects the transceiver mode using the estimated conditions.
5) The customer configures the transceiver with the selected mode.

The advantage of the black box approach is that the optical path can be implemented by the customer alone, thus simplifying the system and almost eliminating the need to work with the carrier. However, the QoT cannot be decomposed into elements such as OSNR_ase and OSNR_nli because the link parameters of the line systems cannot be obtained, so the optimization and fault isolation/location of these elements cannot be performed. This fundamental issue in service operation is difficult to solve with architectural and protocol improvements.

Thus, the remainder of this paper will focus on the DLM approach. We utilize the GNPy open design tool for optical path design [4] with the QoT at exchanged transceiver parameters via the protocol described in Section IV, the known carrier link parameters, and the estimated access link parameters. GNPy assumes a Gaussian noise approximation, which enables optical path design over a wide range of distances, from long-haul to metro, in a very short time and with sufficient accuracy. While [13] has demonstrated QoT maximization with GNPy by collecting open line system (OLS) parameters, we focus on partially unknown OLS parameters.

IV. PROPOSED ARCHITECTURE AND PROTOCOL

The architecture and protocol are designed to resolve the issues discussed in Section III. The left-hand side of Figure 3 shows the proposed architecture. In order to minimize the accuracy degradation in long-distance estimation, we divide the long EtE path into customer and carrier sections. To terminate the customer-side path at the carrier edge, we place an optical switch at the edge and connect to a carrier's measurement receiver. The controller is mounted on the carrier and has a consistency check function and an EtE path design function. The former selects common modes from the mode catalogs of transceivers mounted in A) and B) in the figure. The latter inputs the QoT data and transceiver information estimated by DLM into the optical path design tool and designs the EtE optical path. Then the optical switch reconnects the customer to the destination to establish the EtE path.

The information needed to design the EtE optical path are link parameters in the access and carrier sections, and the capability of customer transceivers. Here, carrier link parameters are generally not disclosed to customers, whereas the configurable modes of the customer transceiver should be accessible to the carrier because it is necessary to manage the quality and safety of the network. The proposed system also allows for the connection between the transceivers of different vendors. Because the configurable modes generally vary by vendor, generation, and version, the carrier requires a mechanism to automatically collect transceiver capability information and perform consistency checks.

Figure 3 shows the overall architecture and protocol including the functional deployment of the DLM approach. A) and B) denote user sites A and B, respectively, and a), b), c), and d) denote the optical switch, measurement equipment, controller, and carrier link, respectively. As shown in Figure 3, all components are handled by the carrier's controller through the secure channel. Transceivers A) and B) are connected to the nearest a) via AALs A and B, respectively. Both switches are connected by d). a) can switch the signal from the AAL to either b) or d). The mode catalog (bitrate, modulation format, FEC type, etc.) is generated and sent to the carrier's controller (see upper right of Figure 3). The controller compares the mode catalog received from A) and B) to create a list of common

OSNR is estimated on the basis of the obtained physical layer parameters. Figure 4 shows the experimental system and estimated PP. Two white-box transponders were prepared for deployment at user sites and connected in the configuration shown in the figure. The white-box transponders are based on the Cassini and Goldstone network operation system (NOS), which are openly specified by TIP, and the TAI, which disaggregates hardware and software in the control of coherent transceivers. Coherent modules from two different vendors are applied, and vendor dependencies are absorbed by the TAI library. This enables transceivers from different vendors to be controlled in a unified manner. The transmission section consists of AALs A and B, which connect the carrier and the user, and the carrier link. AAL A includes 50-km standard single mode fibers (SMFs) and an erbium-doped fiber amplifier modes with which A) and B) can interconnect (see middle right of Figure 3). If no interconnectable mode exists, an error is returned to A) and B). The controller obtains a list of optical path routes from d) in advance and then switches a) to connect A), B), and b), as shown in the lower right of Figure 3. A), B), and b) are test coherent links to estimate the QoT of the bidirectional AAL. Here, functions for estimating QoT are implemented in A), B), and b). All estimated QoT data is collected by the controller, which then calculates the required values for the optical path setup between A) and B) for each optical path route in d) using an optical path design tool. The controller calculates the optimal optical path route and transceiver configuration by referring to the optical path common mode list in d) and sets up and optimizes a) and d) to establish an EtE operational coherent link between A) and B). If no interoperable mode exists, the controller returns an error to A) and B).

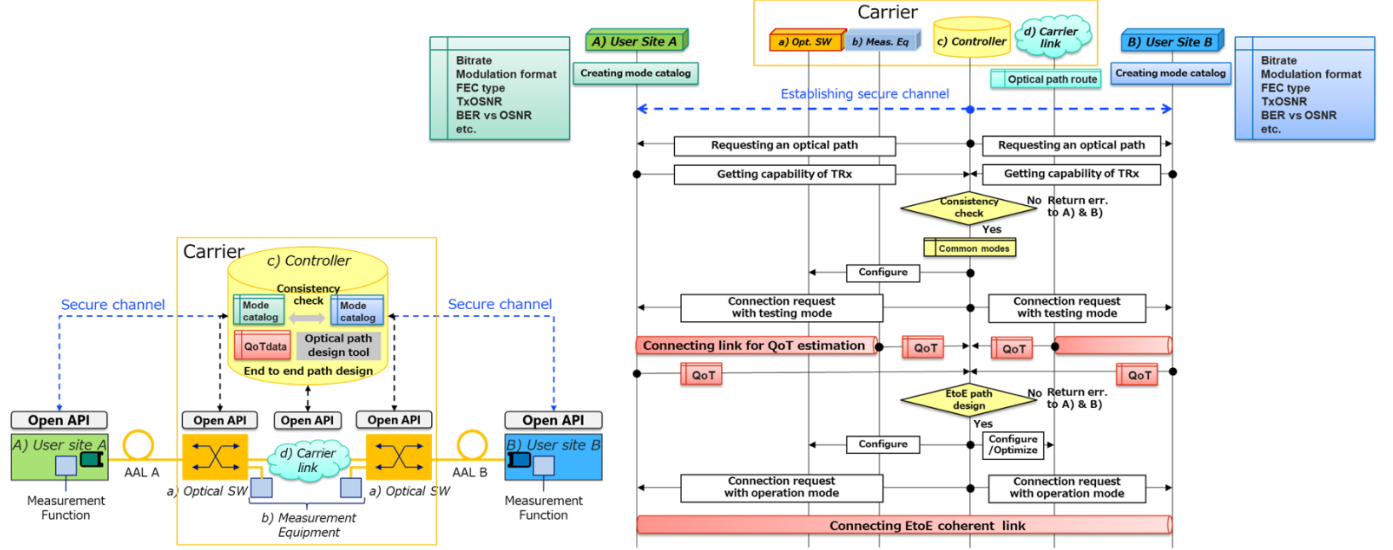

Fig. 3. Proposed architecture and protocol for executing DLM approach.

## V. EVALUATION OF PROPOSED SCHEME

### A. Experimental Test

We experimentally evaluate the proposed scheme to verify whether the optical path can be setup automatically within 10 minutes. First, the PP of the AAL is measured by DLM, and the

(EDFA).

The optical switch can switch the signal from the AAL to the measurement equipment and carrier link. We used the Open ROADM YANG model as reference and developed a common control interface for abstracting multi-vendor switches.

The controller in Figure 4 is connected to white-box transponders and optical switches via control lines and collects capability and QoT data from transceivers and measurement equipment. The controller uses the GNPy to estimate the EtE QoT, determine the optimal configuration with parameters of all links, and then the controller provides instructions for opening the optical path.

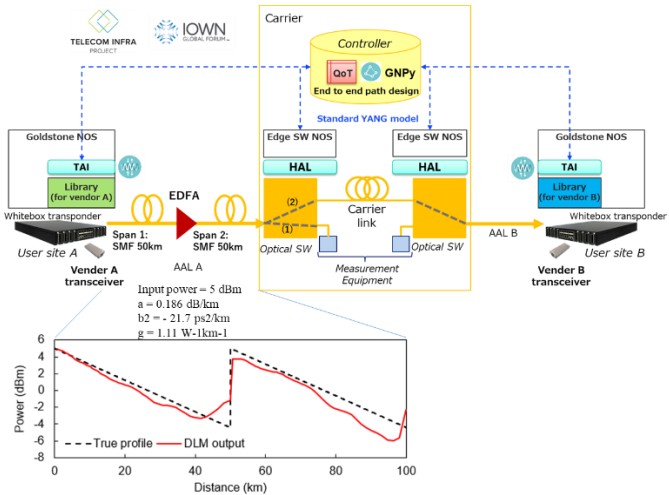

**Fig. 4.** Experimental system and estimated power profile.

First, the PP of AAL A is estimated by DLM by connecting the optical switch as indicated by (1) in Figure 4; the PP is used to estimate the fiber loss/OA gain for the input to GNPy. In this experiment, an offline experimental system is used to perform DLM, but in the future, it will be possible to perform DLM with only a commercially available transceiver. The transmitter is installed on the carrier side, and when DLM starts, the signal is sent to AAL A via an optical switch to monitor the AAL. We used the same offline system as in [2]. The captured waveform on the customer side was sent to a network controller (PC), and demodulation and PPE were performed there. For the PPE, we used the gradient learning of the split-step Fourier method (SSFM) for the Manakov equation [2]. The DLM output of AAL A is shown at the bottom of Figure 4. Note that the vertical axis is the true optical power in dBm. The estimated PP closely matches the true profile. The optical fiber loss, optical fiber input power, and intermediate amplifier gain were estimated from the obtained PP by linear fitting. While the true values were 0.186 dB/km optical fiber loss, 5.0 dBm optical fiber input power, and 9.42 amplifier gain), the estimated values were 0.175 dB/km (1st span) and 0.212 dB/km (2nd span) for optical fiber loss, 4.47 dBm (1st span) and 4.54 dBm (2nd span) for input power, and 8.83 dB for the amplifier gain. For the carrier link, the link parameter is assumed to be known because it is managed by the carrier. For simplicity, only AAL A is used for DLM estimation, but the same method can be used to estimate the transmission quality for the carrier link. In addition, for the transceiver performance parameters (TxOSNR, BER vs. OSNR), we assume that their values are known because they can be obtained from the controller by using the architecture and protocol described in Section IV.

After DLM for the AAL is completed, the switch is changed to (2), and the link parameters of AAL A, B, and the carrier link are input to GNPy. The EtE transmission quality generalized signal to noise ratio (GSNR) was estimated. The transceiver used in this experiment supports 100Gbps dual-polarization (DP) QPSK 32GBd and 200Gbps DP-16QAM 32GBd. In this experiment, the GSNR total was large enough to achieve error-free operation even with 200 Gbps, so the 200 Gbps mode with a high bit rate per optical frequency bandwidth was automatically selected as the optimal configuration. An optical link was established between the transceivers at user sites A and B to verify error-free operation.

Table I summarizes the time required to set up the optical path. The DLM preprocess is the demodulation process for generating inputs to the DLM as shown in Fig. 2.The DLM time is the time taken from the start to the end of iterative learning of the SSFM. The total link parameter estimation took 36 + 27 = 63 sec. to design the optical path using GNPy, and about 73 sec. to boot the transceiver from initial state. In total, the optical path was automatically constructed in about 137 sec.

TABLE I
Time required to set up optical path

|  | Time (sec) |
|---|---|
| **DLM preprocess** | 36 |
| **DLM (Learning of SSFM)** | 27 |
| **Optical path design** | 1 |
| **Transceiver cold start** | 73 |
| **Total** | 137 |

*B. Potential of Margin Reduction*

We have shown how the proposed scheme can estimate the link parameters of an optical path and select the optimal configuration. Because the AAL characteristics can be estimated by DLM, the OSNR margin under the worst-case conditions can be avoided. The OSNR margins that can be reduced by the proposed scheme are the OSNR_ase and the OSNR_nli margins. The former is the margin for power losses caused by an aging OA, connector losses, fusion splicing losses, etc. The latter is the margin for nonlinear noise. In this section, we discuss the extent to which the former power loss margin can be reduced by the proposed scheme.

Here, we define the OSNR margin as the amount of OSNR degradation at the Rx when additional 2-dB losses occur in all spans. The assumptions of the system are shown in the upper part of Figure 5. The fiber input power was uniformly set to 0 dBm for all spans, and the span loss was assumed to be 16 dB under normal conditions (0.2 dB/km, equivalent to 80 km standard) and 18 dB under worst-case conditions when 2 dB of optical power degradation occurs. The gain of the OA was set to compensate for all span loss, and the NF was set to 5.0 dB. The signal was 64 GBd with a wavelength of 1550 nm.

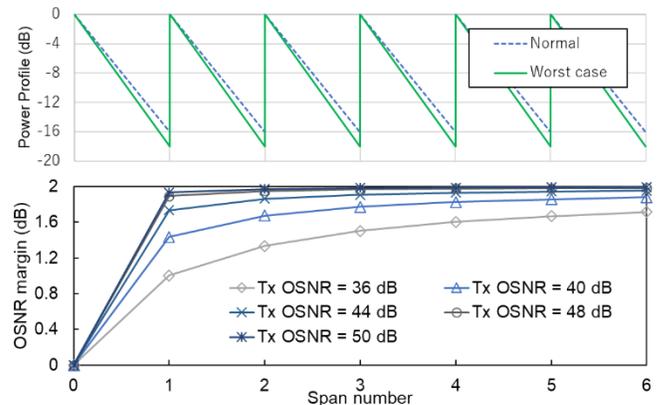

**Fig. 5.** Results of OSNR margin analysis.

The OSNR margins as a function of the total span number are shown in the lower part of Figure 5. Here, Tx OSNR means the OSNR at the output of the coherent transceiver. In a longer distance regime, the effect of the in-line amplifier becomes dominant. In such a regime, an additional span loss of 2 dB always causes an OSNR degradation of nearly 2 dB. In the case of short spans, TxOSNR is the dominant deciding factor for the received OSNR, so the effect of the additional 2-dB span loss is smaller. However, when high TxOSNR is assumed, the effect of ASE noise in the in-line amplifier becomes dominant and the required OSNR margin approaches 2 dB.

Assuming DLM has infinite accuracy, these OSNR margins can be eliminated. The accuracy of the optical power level diagram estimated by DLM is discussed in [2] and [12]. Although the accuracy of DLM at this time is not as high as that of conventional analog measuring instruments such as OTDR, it is still a developing technology, and improvements in accuracy have been discussed in [12]. The above OSNR margins can be reduced in the future when DLM can achieve accuracy comparable to that of OTDRs.

## VI. REMAINING RESEARCH ISSUES

While the DLM approach offers various advantages, the challenge is developing measurement equipment and optical switches in the carrier network at low cost. In our experiment, a single wavelength was used for simplicity, but it is necessary to examine the effects of the wavelength dependence of OAs and fiber nonlinear effects when deploying DLM in WDM. It is also necessary to solve problems related to accommodation design, such as available wavelength assignment in the carrier link and optical path route design, which are being discussed at the IOWN Global Forum [14].

The number of transceiver modes has been increasing as DSP LSIs evolve. There is a need for a method of efficiently selecting the optimal mode without measuring the QoT characteristics for each mode [4]. A method of selecting the optimal one without trying all of them, which is only a selection of the modulation format, has been studied in [6]. This method also requires the BER-OSNR characteristics of the customer's transceiver, but there is no open API to obtain them. If the transceiver characteristics can be obtained, it should be possible to design optical paths that take into account the individual differences of transceivers, which is expected to reduce margins or improve frequency utilization efficiency.

Regarding the confidentiality of the link parameters between the customer and the carrier, in this paper, we assumed that the carrier obtains the estimated link parameters, but the customer may not allow the carrier to obtain the parameters. In virtual networks, cryptographic techniques such as secure multi-party computation have been proposed to set up virtual paths while keeping the information secret [15]. A similar approach may be used in optical transmission networks to select optimal transceiver mode without revealing the link parameters of the optical path.

As described in Section I, optical path setting requests may be concentrated on carriers as the frequency of optical transmission network usage and the number of optical lines increase. Such concentrated requests are often seen in mobile networks, where admission control has been introduced. However, as seen in our experiments, optical path setting takes more time than mobile terminal registration, so it cannot be used as is for optical path setting. In particular, at the time of a large-scale failure due to a disaster, not only do all terminals request re-connection again upon recovery, but the estimation process cannot be omitted because the link parameters of the optical path may have changed due to the disaster. Thus, there is a need for a new admission control that is optimized for the optical path setting.

## VII. CONCLUSION

We proposed an architecture and protocol for cooperative optical path design between carriers and customers. The protocol utilizes DLM, a technique for estimating optical link parameters, to automatically configure the optical path with optimal QoT between customer-owned transceivers connected via AAL whose link parameters are unknown and carrier-managed networks. DLM enables estimation of $OSNR\_ase$ and $OSNR\_nli$ and reduces the margin for receiver-end OSNR degradation by up to 2 dB for each span, assuming a 2 dB increase in span loss. Our experiments verified that the proposed architecture and protocol enable the customer to optimize and establish an EtE optical path including AALs in about 137 seconds.

## REFERENCES

[1] M. Filer et al., "Low-margin optical networking at cloud scale," *Journal of optical communications and networking,* vol. 11, issue 10, pp. C94-C108, 2019.
[2] T. Sasai et al., "Digital longitudinal monitoring of optical fiber communication link," *Journal of lightwave technol.,* vol. 40, no. 8, Apr. 2022.
[3] T. Tanimura, S. Yoshida, K. Tajima, S. Oda, and T. Hoshida, "Concept and implementation study of advanced DSP-based fiber-longitudinal optical power profile monitoring toward optical network tomography," *Journal of optical communications and networking,* vol. 13, no. 10, Oct. 2021.
[4] V. Curri, "GNPy model of the physical layer for open and disaggregated optical networking," *Journal of optical communications and networking,* vol. 14, issue 6, pp. C92-C104, 2022.
[5] K. Kaeval et al., "QoT assessment of the optical spectrum as a service in disaggregated network scenarios," *Journal of optical communications and networking,* vol. 13, no. 10, Oct. 2021.
[6] K. Anazawa et al., "Automatic modulation-format selection with white-box transponders: design and field trial," *OECC 2021,* paper JS2A.14.
[7] A. Gouin et al., "Real-time optical transponder prototype with autonegotiation protocol for software defined networks," *Journal of optical communications and networking,* vol. 13, no. 9, Sep. 2021.
[8] L. Alahdab et al., "Alien wavelengths over optical transport networks," *J. opt. commun. netw.*, vol. 10, no. 11 Nov. 2018.
[9] L. Liu et al., "Field and lab trials of PCE-based OSNR-aware dynamic restoration in multi-domain GMPLS-enabled translucent WSON," *Optics express,* vol. 19, no. 27 Dec. 2011.
[10] A. M. Fagertun, B. Skjoldstrup, "Flexible transport network expansion via open WDM interface," *international conference on computing, networking and communications, optical and grid networking symposium,* 2013.
[11] T. Tanimura, S. Yoshida, K. Tajima, S. Oda, and T. Hoshida, "Fiber-longitudinal anomaly position identification over multi-span transmission link out of receiver-end signals," *J. lightwave technol.,* vol. 38, issue 9, pp. 2726-2733, 2020.
[12] T. Sasai, M. Nakamura, E. Yamazaki, and Y. Kisaka, "Precise longitudinal power monitoring over 2,080km enabled by step size selection of split step fourier method," *OFC 2022*, paper Th1C.4.
[13] G. Borraccini, "Using QoT for open line controlling and modulation format deployment: an experimental proof of concept," *ECOC 2020.*
[14] IOWN global forum technical paper, "Open ALL-Photonic network functional architecture 1.0," Jan. 2022.
[15] T. Mano et al., "Efficient virtual network optimization across multiple domains without revealing private information," *IEEE transactions on network and service management,* vol. 13, issue 3, Sep. 2016.

BIOGRAPHIES

Hideki Nishizawa received B.E. and M.E. degrees in physics from Chiba University, Chiba, Japan, in 1994 and 1996, respectively. In 1996, he joined NTT Laboratories, Japan, where he has been engaged in research on open and disaggregated optical systems.

Takeo Sasai received B.E. and M.E. degrees in electrical engineering from the University of Tokyo, Japan, in 2014 and 2016, respectively. In 2016, he joined NTT Laboratories, Japan, where he has been engaged in the digital longitudinal monitoring of optical fiber link.

Takeru Inoue received his Ph.D. in information science from Kyoto University, Japan, in 2006. He joined NTT Laboratories in 2000 and currently serves as a Distinguished Researcher.

Kazuya Anazawa received his B.E. and M.E. degrees from the University of Aizu in 2016 and 2018, respectively. He joined NTT Laboratories in 2018. His current research interests are autonomous control of optical transport networks.

Toru Mano received B.E. and M.E. degrees from the University of Tokyo in 2009 and 2011, respectively. He joined NTT Labs in 2011. He received the Ph.D. degree in computer science and information technology from the Hokkaido University in 2020. His research interests are network architectures, network optimization, softwarization of networking.

Kei Kitamura received B.E., M.E. degrees in information and communication engineering from the University of Electro-Communications, Tokyo, Japan, in 2008 and 2010, respectively. In 2010, he joined NTT Laboratories and has been engaged in research on network interfaces and large-capacity transmission systems.

Yoshiaki Sone received his M.E. degree in electronics engineering from Tohoku University, Miyagi, in 2003. In 2003, he joined NTT Laboratories to focus his research on network engineering technologies for optical transport networks.

Tetsuro Inui received his B.E. degree in electric engineering from Waseda University, Tokyo, Japan in 1995 and M.E. degree in electronic engineering from the University of Tokyo, in 1998. In 1998, he joined NTT Laboratories, Japan, where he has been engaged in research on high-speed optical transport systems and optical transport network architecture.

Koichi Takasugi received the B.E. degree in computer science from Tokyo Institute of Technology in 1995, the M.E. degree from Japan Advanced Institute of Science and Technology in 1997 and Ph.D. in engineering from Waseda University in 2004. In 1997, he joined NTT Laboratories, Japan, where he has been engaged in research on wireless and optical transport networks.